\documentstyle[aps,prl,multicol,epsfig]{revtex}
\newcommand{\bq}{\begin{equation}}
\newcommand{\eq}{\end{equation}}
\newcommand{\bqa}{\begin{eqnarray}}
\newcommand{\eqa}{\end{eqnarray}}
\newcommand{\nn}{\nonumber \\}

\begin{document}
\draft 
\title{Holon Pair Bose Condensation in High $T_c$ Cuprates;  Symmetry Breaking and Supersymmetry Conditions}
\author{Sung-Sik Lee and Sung-Ho Suck Salk$^a$}
\address{Department of Physics, Pohang University of Science and Technology,\\
Pohang, Kyoungbuk, Korea 790-784\\
$^a$ Korea Institute of Advanced Studies, Seoul 130-012, Korea\\}
\date{\today}

\maketitle

\begin{abstract}
Using the t-J Hamiltonian of U(1) slave-boson symmetry, bose condensation is discussed by showing the occurrence of symmetry breaking in the hole doped high $T_c$ cuprates.
The symmetry breaking is shown to take place with the d-wave hole pairing, but not with the s-wave hole pairing.
Based on a derived supersymmetry Hamiltonian we find that there exists a possibility of supersymmetry conditions in association with the pairing order parameters of both spinon and holon.
\end{abstract}
\begin{multicols}{2}

Using the U(1) slave-boson approach\cite{KOTLIAR}-\cite{UBBENS} to the t-J Hamiltonian, previous studies showed a possibility of d-wave superconductivity(bose condensation) referring to a composite of d-wave spinon(fermion) pairing and single holon(boson) condensation.
Another possibility is the d-wave superconductivity associated with the composite of d-wave spinon pairing, $< f_{i \uparrow} f_{j \downarrow} >$ and s-wave holon pairing, $< b_i^{\dagger} b_j^{\dagger}>$ for the hole-pairing order parameter of $\Delta \approx < c_{i \uparrow} c_{j \downarrow} > = < f_{i \uparrow} f_{j \downarrow} > < b_i^{\dagger} b_j^{\dagger}>$.
Lately Wen and Lee proposed an SU(2) slave-boson theory and questioned whether the single holon bose condensation or the holon pair bose condensation is favored in the superconducting phase\cite{WEN}.
In the present study, by introducing an improved approach of treating the Heisenberg term over our earlier theory\cite{GIMM}, we discuss bose condensation by showing the occurrence of symmetry breaking with respect to holon pairing.
In low dimensions, the spontaneous ordering of infinite-range can not occur at finite temperatures\cite{HMW}.
Thus we stress that the term 'bose condensation' refers to a finite range order, but not the infinite range order, as is often alluded in the literatures\cite{KOTLIAR}-\cite{WEN}.
In addition, we explore a possible existence of the supersymmetry conditions in association with the pairing order parameters of both the spinon(fermion) and the holon(boson) with both the spinon and holon pairing order parameters.

We write the t-J Hamiltonian, 
%%%%%%%%%%%%%%TWO COLUMN%%%%%%%%%%%%%%%%%%%%%%%%%%%
\begin{eqnarray}
H & = & -t\sum_{<i,j>}(c_{i\sigma}^{\dagger}c_{j\sigma} + c.c.) + J\sum_{<i,j>}(
{\bf S}_{i} \cdot {\bf S}_{j} - \frac{1}{4}n_{i}n_{j}) \nn
&& - \mu_{0}\sum_{i,\sigma} c_{i\sigma}^{\dagger}c_{i\sigma},
\label{eq:tjmodel1}
\end{eqnarray}
%%%%%%%%%%%%%%%%%%%%%%%%%%%%%%%%%%%%%%%%%%%%%%%%%%%
where $ {\bf S}_{i} \cdot {\bf S}_{j} - \frac{1}{4}n_{i}n_{j} =  -\frac{1}{2} ( c_{i\downarrow}^{\dagger}c_{j\uparrow}^{\dagger}-c_{i\uparrow}^{\dagger}c_{j\downarrow}^{\dagger}) (c_{j\uparrow}c_{i\downarrow}-c_{j\downarrow} c_{i\uparrow})$.
${\bf S}_{i}$ is the electron spin operator at site $i$, ${\bf S}_{i}=\frac{1}{2}c_{i\alpha}^{\dagger} \bbox{\sigma}_{\alpha \beta}c_{i\beta}$ with $\bbox{\sigma}_{\alpha \beta}$, the Pauli spin matrix element. $n_i$ is the electron number operator at site $i$, $n_i=c_{i\sigma}^{\dagger}c_{i\sigma}$.
Allowing the occupancy constraint, $c_{i\sigma}^\dagger c_{i\sigma} \leq 1$, the U(1) slave-boson representation of the above t-J Hamiltonian is,
%%%%%%%%%%%%%%TWO COLUMN%%%%%%%%%%%%%%%
\begin{eqnarray}
H & = & -t\sum_{<i,j>}(f_{i\sigma}^{\dagger}f_{j\sigma}b_{j}^{\dagger}b_{i} + c.c.) \nonumber \\
&& -\frac{J}{2} \sum_{<i,j>} b_i b_j b_j^{\dagger}b_i^{\dagger} (f_{i\downarrow}^{\dagger}f_{j\uparrow}^{\dagger}-f_{i\uparrow}^ {\dagger}f_{j\downarrow}^{\dagger})(f_{j\uparrow}f_{i\downarrow}-f_{j\downarrow} f_{i\uparrow}) \nn
&& - \mu_{0}\sum_{i,\sigma} f_{i\sigma}^{\dagger}f_{i\sigma}   + i\sum_{i} \lambda_{i}(f_{i\sigma}^{\dagger}f_{i\sigma}+b_{i}^{\dagger}b_{i} -1),
\label{eq:tjmodel}
\end{eqnarray}
%%%%%%%%%%%%%%%%%%%%%%%%%%%%%%%%%%%%%
where $f_{i\sigma}(f_{i\sigma}^{\dagger})$ is the spinon annihilation(creation) operator and $b_{i}(b_{i}^{\dagger})$, the holon annihilation(creation) operator.
Here the holon($b$) represents the spinless boson with charge $+e$ and the spinon($f$), the chargeless fermion with spin $\frac{1}{2}$.
$\lambda_{i}$ is the Lagrangian multiplier to enforce the local single occupancy constraint.
It is of note that $b_i b_i^\dagger b_j b_j^\dagger = 1$ was assumed in the previous studies\cite{KOTLIAR}-\cite{GIMM}.
This identity is valid only for the case of no charge fluctuations at each site which can occur only at half-filling(i.e., no hole doping) where site to site hopping is prohibited.
For generality we allow such fluctuations for the quantum systems of interest in the hole doped systems.

We cast the Heisenberg coupling term(second term of Eq.(\ref{eq:tjmodel})) into  
%%%%%%%%%%%%%%TWO COLUMN%%%%%%%%%%%%%%%
\bqa
\lefteqn{-\frac{J}{2} b_i b_j b_j^{\dagger}b_i^{\dagger} (f_{i\downarrow}^{\dagger}f_{j\uparrow}^{\dagger}-f_{i\uparrow}^ {\dagger}f_{j\downarrow}^{\dagger})(f_{j\uparrow}f_{i\downarrow}-f_{j\downarrow} f_{i\uparrow}) } \nn
& = & -\frac{J}{2} \Bigl< b_i b_j b_j^{\dagger}b_i^{\dagger} \Bigr> (f_{i\downarrow}^{\dagger}f_{j\uparrow}^{\dagger}-f_{i\uparrow}^ {\dagger}f_{j\downarrow}^{\dagger})(f_{j\uparrow}f_{i\downarrow}-f_{j\downarrow} f_{i\uparrow}) \nn
& & -\frac{J}{2} b_i b_j b_j^{\dagger}b_i^{\dagger} \Bigl< (f_{i\downarrow}^{\dagger}f_{j\uparrow}^{\dagger}-f_{i\uparrow}^ {\dagger}f_{j\downarrow}^{\dagger})(f_{j\uparrow}f_{i\downarrow}-f_{j\downarrow} f_{i\uparrow}) \Bigr> \nn
& & + \frac{J}{2} \Bigl< b_i b_j b_j^{\dagger}b_i^{\dagger} \Bigr> \Bigl< (f_{i\downarrow}^{\dagger}f_{j\uparrow}^{\dagger}-f_{i\uparrow}^ {\dagger}f_{j\downarrow}^{\dagger})(f_{j\uparrow}f_{i\downarrow}-f_{j\downarrow} f_{i\uparrow}) \Bigr> \nn
& & -\frac{J}{2} \Bigl( b_i b_j b_j^{\dagger}b_i^{\dagger} - \Bigl< b_i b_j b_j^{\dagger}b_i^{\dagger}\Bigr> \Bigr) \times \nn
&& \Bigl( (f_{i\downarrow}^{\dagger}f_{j\uparrow}^{\dagger}-f_{i\uparrow}^ {\dagger}f_{j\downarrow}^{\dagger} ) ( f_{j\uparrow}f_{i\downarrow}-f_{j\downarrow} f_{i\uparrow}) \nn
&& - \Bigl<(f_{i\downarrow}^{\dagger}f_{j\uparrow}^{\dagger}-f_{i\uparrow}^ {\dagger}f_{j\downarrow}^{\dagger})(f_{j\uparrow}f_{i\downarrow}-f_{j\downarrow} f_{i\uparrow})\Bigr> \Bigr), \label{eq:mf_fluc} \nn
\vspace{-0.5cm}
\eqa
%%%%%%%%%%%%%%%%%%%%%%%%%%%%%%%%%%%%%%%
where the first three terms represent the mean field contributions and the last term involves correlations between the fluctuations of the pairing order parameters of the spinon and the holon respectively.
The contribution of the last term may be minimal particularly in the underdoped region where there exists a large difference between the pseudogap(spin gap) transition temperature and the superconducting transition temperature.

By introducing the Hubbard-Stratonovich fields, ${\rho}_{i}^{k}$, $\chi_{ji}$ and $\Delta_{ji}^{f}$ corresponding to the spinon channels involving the direct, exchange, and pairing interactions in the first term of Eq.(\ref{eq:mf_fluc}), we obtain the effective Hamiltonian from Eq.(\ref{eq:tjmodel})\cite{1_DELTA},
%%%%%%%%%%%%%%TWO COLUMN%%%%%%%%%%%%%%%
\begin{eqnarray}
\lefteqn{ H_{eff}   =  J(1-\delta^2) \sum_{<i,j>} \Bigl[ |\Delta_{ji}^{f}|^{2} + \frac{1}{4}|\chi_{ji}|^{2} }\nn
 && - \frac{1}{4} \chi_{ji}^{*} \Bigl( f_{j\sigma}^{\dagger}f_{i\sigma}+\frac{4t}{J(1-\delta)^2}b_{j}^{\dagger}b_{i} \Bigr) -c.c. \nn
 && + \frac{1}{4} n_{i} - \frac{1}{2} \Delta_{ji}^{f*}(f_{j\uparrow}f_{i\downarrow}-f_{j\downarrow}f_{i\uparrow})-c.c. \Bigr] \nn
 && + \frac{J(1-\delta)^2}{2} \sum_{<i,j>}\sum_{k=0}^3 \Bigl( ({\rho}_{i}^{k})^{2 } - (\rho_{i}^{k})(f_{j}^{\dagger}\sigma^{k}f_{j}) \Bigr)  \nn
 && -\mu_{0}\sum_{i} f_{i\sigma}^{\dagger}f_{i\sigma}   -i\sum_{i}\lambda_{i}(f_{i\sigma}^{\dagger}f_{i\sigma}+b_{i}^{\dagger}b_{i}-1) \nn
  && - \sum_{<i,j>} \frac{J}{2}|\Delta^f_{ij}|^2 b_{i}b_{j}b_{i}^\dagger b_{j}^\dagger  +  \frac{4t^{2}}{J(1-\delta)^2}\sum_{<i,j>}(b_{j}^{\dagger}b_{i})(b_{i}^{\dagger}b_{j}),
  \label{eq:mf_hamiltonian1}
  \end{eqnarray}
%%%%%%%%%%%%%%%%%%%%%%%%%%%%%%%%%%%%%%%%
where $|\Delta^f_{ij}|^2 = \Bigl< (f_{i\downarrow}^{\dagger}f_{j\uparrow}^{\dagger}-f_{i\uparrow}^ {\dagger}f_{j\downarrow}^{\dagger})(f_{j\uparrow}f_{i\downarrow}-f_{j\downarrow} f_{i\uparrow}) \Bigr>$ for spinon pairing.
In the above equation we ignored the correlation effects between the fluctuations of the order parameters.
$\sigma$ is the Pauli spin matrices and $\delta$, the hole doping rate.
The parenthesis $(b_{j}^{\dagger}b_{i})(b_{i}^{\dagger}b_{j})$ in the last term represents only the exchange interaction channel. 
The exchange channel involves a large repulsive(positive) energy of order $U \approx \frac{4t^2}{J}$ and has been ignored\cite{UBBENS}\cite{WEN}.
The holon-holon interaction term(one before the last term) involves holon pairing which is coupled with(varies with) the spinon pairing order $\Delta^f$.
The Lagrangian multiplier term will be incorporated into the effective chemical potential terms of the spinon and the holon respectively.

The Hubbard Stratonovich transformation for the hole pairing channel leads to
%%%%%%%%%%%%%%TWO COLUMN%%%%%%%%%%%%%%%
\bqa
\displaystyle
\lefteqn{e^{\sum_{<i,j>} \frac{J}{2}|\Delta^f_{ij}|^2 b_{i}^\dagger b_{j}^\dagger b_{i} b_{j}}  \propto } \nn
&& \int \prod_{<i,j>} d\Delta_{ji}^{b*} d\Delta_{ji}^{b} e^{ -\sum_{<i,j>} \frac{J}{2}|\Delta^f_{ij}|^2 \Bigl[ |\Delta_{ji}^{b}|^{2} - \Delta_{ji}^{b*} (b_{j}b_{i}) - c.c \Bigr] },
\label{eq:holon_Hubbard_Stratonovich}
\eqa
%%%%%%%%%%%%%%%%%%%%%%%%%%%%%%%%%%%%%%%
where $\Delta_b$ is the scalar boson field of holon pairing.
Using Eq.(\ref{eq:holon_Hubbard_Stratonovich}) above and the saddle point approximation we obtain the mean field Hamiltonian from Eq.(\ref{eq:mf_hamiltonian1}),
%%%%%%%%%%%%%%TWO COLUMN%%%%%%%%%%%%%%%
\begin{eqnarray}
\lefteqn{H^{MF}=\sum_{<i,j>}\Bigl[ \frac{J(1-\delta)^2}{2} |\Delta_{ji}^{f}|^{2} + \frac{J(1-\delta)^2}{4} |\chi_{ji}|^{2} + \frac{J}{2}|\Delta^f_{ij}|^2|\Delta_{ji}^{b}|^{2} }\nn
&&+\frac{J}{2} |\Delta^f_{ji}|^2 \delta^2 \Bigr]    -\frac{J(1-\delta)^2}{2} \sum_{<i,j>} \Bigl[ \Delta_{ji}^{f*} (f_{j\uparrow}f_{i\downarrow}-f_{j\downarrow}f_{i\uparrow}) \nn
 &&+ c.c. \Bigr]   -\frac{J(1-\delta)^2}{4} \sum_{<i,j>} \Bigl[ \chi_{ji}^{*} (f_{j\sigma}^{\dagger}f_{i\sigma}) + c.c. \Bigr] + \nonumber \\
&& + \frac{J(1-\delta)^2}{2}\sum_{<i,j>}\sum_{k=0}^3 \Bigl( ({\rho}_{i}^{k})^{2} - \sum_{k=0}^{3} (\rho_{i}^{k})(f_{j}^{\dagger}\sigma^{k}f_{j}) \Bigr) \nn
 & & -t \sum_{<i,j>} \Bigl[ \chi_{ji}^{*}(b_{j}^{\dagger}b_{i}) + c.c.  \Bigr] -\sum_{<i,j>} \frac{J}{2}|\Delta^f_{ij}|^2 \Bigl[ \Delta_{ji}^{b*} (b_{i}b_{j}) + c.c. \Bigr] \nn
 && - \sum_{i,\sigma} \mu^{f}_{i} \left( f_{i\sigma}^{\dagger} f_{i\sigma} -(1-\delta) \right)  -\sum_{i} \mu_{i}^{b} ( b_{i}^{\dagger}b_{i} -\delta ),
 \label{eq:mf_hamiltonian2}
 \end{eqnarray}
%%%%%%%%%%%%%%%%%%%%%%%%%%%%%%%%%%%%%%%
where $\chi_{ji}= < f_{j\sigma}^{\dagger}f_{i\sigma} + \frac{4t}{J(1-\delta)^2} b_{j}^{\dagger}b_{i}>= \chi_{ji}^f + \frac{4t}{J(1-\delta)^2} \chi_{ji}^b$ with $\chi_{ji}^f  = < f_{j\sigma}^{\dagger}f_{i\sigma}>$ and $\chi_{ji}^b = < b_{j}^{\dagger}b_{i}>$, $\Delta_{ji}^{f}=< f_{j\uparrow}f_{i\downarrow}-f_{j\downarrow}f_{i\uparrow} >$, $\Delta_{ji}^{b} = <b_{j}b_{i}>$, $\rho_{i}^{k}=<\frac{1}{2}f_{i}^\dagger \sigma^k f_i>$,  $\mu_{i}^{f}=\mu_{0} + i\lambda_{i}-J(1-\delta)^2/2$ and $\mu_{i}^{b}=i\lambda_{i} - \frac{J}{2} \sum_{i=i \pm \hat x, \hat y} |\Delta_{ji}^{f}|^{2}$.
$\rho_{i}^{k=1,2,3}=<S_i^k>$ will be taken to be $0$\cite{UBBENS} and $\rho_i^{k=0}=\frac{1}{2}<n_i>$ will be incorporated into the spinon chemical potential term.
For simplicity we allow uniform(site-independent) chemical potentials, $\mu^{f}_{i}=\mu^{f}$ and $\mu^{b}_{i}=\mu^{b}$.

We now introduce a uniform(that is, site-independent) hopping order parameter with the flux phase, $\chi_{ji}=\chi e^{\pm i\theta }$, where the sign $+(-)$ is for the counterclockwise(clockwise) direction around a plaquette and the pairing order parameters, $ \Delta_{ji}^{f}=\Delta_f e^{\pm i\tau^{f}} \mbox{ and } \Delta_{ji}^{b}=\Delta_b e^{\pm i\tau^{b}}$, where the sign $+(-)$ is for the ${\bf ij}$ link parallel to $\hat x$ ($\hat y$) and $\Delta_b$, $\Delta_f$ and $\chi$ denote uniform(site independent) amplitudes\cite{UBBENS}.
The subscripts $i$ and $j$ will be deleted from now on.
We employ the Bogoliubov-Valatin transformation following the momentum space representation of the mean field Hamiltonian $H^{MF}$ in Eq.(\ref{eq:mf_hamiltonian2}). 
The resulting Hamiltonian is diagonalized as
%%%%%%%%%%%%%%TWO COLUMN%%%%%%%%%%%%%%%
\begin{eqnarray}
\lefteqn{ H^{MF}  =  N (1-\delta)^2 J \Bigl( \frac{\chi^2}{2} +  \Delta_f^{2} \Bigr) } \nn
&& + \sum_{k, s=\pm 1}^{'} E_{ks}^{f}(\alpha_{ks}^{\dagger}\alpha_{ks} - \beta_{ks}\beta_{ks}^{\dagger}) + NJ\Delta_f^2 ( \Delta_b^{2} + \delta^2 ) \nn
 &&+ \sum_{k, s=\pm 1}^{'} \left( E_{ks}^{b} h_{ks}^{\dagger} h_{ks} + \frac{E_{ks}^{b} + \mu^b}{2}) \right)  - N \delta \mu^{f} + N \delta \mu^{b},
\label{eq:diagonalized_hamiltonian}
\end{eqnarray}
%%%%%%%%%%%%%%%%%%%%%%%%%%%%%%%%%%%%%%%
where $\sum^{'}$ denotes the summation over momentum $k$ in the half reduced Brillouin zone, and $s= +1$ and $-1$ represent the upper and lower energy bands of quasiparticles respectively. 

Here $E_{ks}^{f}$ and $E_{ks}^{b}$ are the quasiparticle energies,
\begin{equation}
E_{ks}^{f}  =  \sqrt{(\epsilon_{ks}^{f}-\mu^{f})^{2} + \Bigl( J(1-\delta)^2 \xi_{k}(\tau^{f}) \Delta_f \Bigr)^{2}}, 
\label{eq:spinon_energy}
\end{equation}
for spinons and
\begin{equation}
E_{ks}^{b}  =  \sqrt{(\epsilon_{ks}^{b}-\mu^{b})^{2} - \Bigl( J\Delta_f^2 \xi_{k}(\tau^{b}) \Delta_b \Bigr)^{2}}, 
\label{eq:holon_energy}
\end{equation}
for holons, where
the symbol definitions are, for $\phi=\theta$, $\tau^{f}$ or $\tau^{b}$
\begin{eqnarray}
\xi_{k}(\phi) & = & \sqrt{ \gamma_{k}^{2} \cos^{2} \phi + \varphi_{k}^{2} \sin^{2} \phi }, \label{eq:xi},  \\
\epsilon_{ks}^{f} & = & \frac{J(1-\delta)^2}{2}s\chi \xi_{k}(\theta), \\
\epsilon_{ks}^{b} & = & 2ts \chi \xi_{k}(\theta), 
\end{eqnarray}
with $\phi=\theta$, $\tau^{f}$ or $\tau^{b}$, $ \gamma_{k} = (\cos k_{x} + \cos k_{y})$ and $\varphi_{k} = ( \cos k_{x} - \cos k_{y})$.
Here $\epsilon_{ks}^{f}$ and $\epsilon_{ks}^{b}$ are the quasiparticle energies for spinons and holons respectively in the absence of both the spinon and holon pairing, i.e., $\Delta_f=\Delta_b=0$.
$\alpha_{ks}( \alpha_{ks}^{\dagger})$ and $\beta_{ks}(\beta_{ks}^{\dagger})$ are the annihilation(creation) operators of spinon quasiparticles or 'quasi-spinons' of spin up and spin down respectively, and $h_{ks}(h_{ks}^{\dagger})$, the annihilation(creation) operators of holon quasiparticles or 'quasi-holons'.

Considering a case of equivalence between the quasi-spinon energy and the quasi-holon energy, i.e., $E^f_{ks}=E^b_{ks}$, we obtain from Eq.(\ref{eq:diagonalized_hamiltonian}) 
\begin{equation}
H_{SUSY} = \sum_{k, s=\pm 1}^{'} E_{ks}(g_{ks}^{\dagger}g_{ks} + h_{ks}^{\dagger} h_{ks} ),
\label{eq:susy_hamiltonian}
\end{equation}
with $g_{ks}=\alpha_{ks}$ or $\beta_{ks}$.
The above Hamiltonian is a SUSY(supersymmetry) Hamiltonian; the SUSY algebra\cite{LAHIRI} $ \{ Q,Q \}  =  H_{SUSY}$ is satisfied with the supercharge operator, $ Q  = \sum_{k, s=\pm 1}^{'} \sqrt{\frac{E_{ks}}{2}} \Bigl( g_{ks}^{\dagger} h_{ks} + h_{ks}^{\dagger} g_{ks} \Bigr)$.
Allowing the equality of $E_{ks}^f = E_{ks}^b$ between Eqs. (\ref{eq:spinon_energy}) and (\ref{eq:holon_energy}), we obtain the SUSY conditions of $ \Delta_f  = \Delta^b = 0$, $\chi^f  = \chi^b=  0$ and  $\mu^{f}  =  \mu^{b}$.
These SUSY conditions may be satisfied in the intermediate doping region between the antiferromagnetic phase and the superconducting phase at $T=0K$.
Obviously one of the SUSY conditions, $\Delta_b=0$ is satisfied in this region.
It will be of great interest to experimentally verify the disappearance of the spinon(spin singlet) pairing which may occur as a result of quantum fluctuations at $0 K$.

From the diagonalized Hamiltonian Eq.(\ref{eq:diagonalized_hamiltonian}), we obtain the total free energy, 
%%%%%%%%%%%%%%TWO COLUMN%%%%%%%%%%%%%%%
\begin{eqnarray}
\lefteqn{F=NJ(1-\delta)^2 \Bigl( \Delta_f^{2} + \frac{1}{2}\chi^{2} \Bigr)   } \nn
&& - 2k_{B}T \sum_{k,s=\pm 1}^{'} ln [ \cosh (\beta E_{ks}^{f}/2) ]  - N\delta \mu^{f} - 2Nk_{B}Tln2 \nn
&& + NJ\Delta_f^2 ( \Delta_b^{2} + \delta^2 )+ k_{B}T \sum_{k,s=\pm 1}^{'} ln [1 - e^{-\beta E_{ks}^{b}}] \nn
&& + \sum_{k,s=\pm 1}^{'} \frac{ E_{ks}^{b}+\mu^{b}}{2} + N \delta \mu^{b}.
\label{eq:free_energy}
\end{eqnarray}
%%%%%%%%%%%%%%%%%%%%%%%%%%%%%%%%%%%%%%%
By minimizing the above free energy, the order parameters $\chi$, $\Delta_f$ and $\Delta_b$ are numerically obtained as a function of temperature and doping rate.
The chemical potential of quasi-holons for the hole-doped systems at finite temperature can be determined from the relation $\frac{\partial F}{\partial \mu^{b}} = 0$.
Likewise, the chemical potential of the quasi-spinons can be obtained from $\frac{\partial F}{\partial \mu^{f}} = 0$.

Using Eq.(\ref{eq:free_energy}), we obtain  
%%%%%%%%%%%%%%TWO COLUMN%%%%%%%%%%%%%%%
\begin{eqnarray}
-\frac{\partial F}{\partial \mu^{b}} &=& \sum_{k,s=\pm 1}^{'} \Bigl[ \frac{1}{e^{\beta E_{ks}^{b}}-1} \frac{\epsilon_{ks}^{b}-\mu^{b}}{E_{ks}^{b}} + \frac{\epsilon_{ks}^{b}-\mu^{b}-E_{ks}^{b}}{2E_{ks}^{b}} \Bigr] - N \delta \nn
& = & 0 \label{eq:d_mu_F}, \\
\frac{\partial F}{\partial \Delta_b} &=& 2 J\Delta_f^2 \Delta_b \Bigl[ N - \sum_{k,s=\pm 1}^{'} \Bigl( \frac{1}{e^{\beta E_{ks}^{b}}-1}+ \frac{1}{2} \Bigr) \frac{J\Delta_f^2 \xi_{k}(\tau^{b})^{2}}{2E_{ks}^{b}} \Bigr] \nn
& = & 0 \label{eq:d_delta_F}.
\end{eqnarray}
%%%%%%%%%%%%%%%%%%%%%%%%%%%%%%%%%%%%%%%
It is gratifying to note that in the absence of holon pairing, i.e., $\Delta_b=0$, Eq.(\ref{eq:d_mu_F}) leads to the Bose-Einstein statistics relation $\sum_{k,s=\pm 1}^{'} \frac{1}{e^{\beta (\epsilon_{ks}^{b}-\mu^{b})}-1} = N \delta$ for free holons(quasi-holons).
The free energy of the holon-pair boson is from Eq.(\ref{eq:free_energy}), 
%%%%%%%%%%%%%%TWO COLUMN%%%%%%%%%%%%%%%
\bqa
F^{b} & =  & NJ\Delta_f^2 (\Delta_b^{2} +\delta^2) + k_{B}T \sum_{k,s=\pm 1}^{'} ln [1 - e^{-\beta E_{ks}^{b}}] \nn
&& + \sum_{k,s=\pm 1}^{'} \frac{ E_{ks}^{b}+\mu^{b}}{2} + N\delta\mu^{b}.
\label{eq:holon_free_energy}
\eqa
%%%%%%%%%%%%%%%%%%%%%%%%%%%%%%%%%%%%%%%
This free energy is an even function of the complex holon pair order parameter, $\Delta_{ji}^{b} = \Delta_b e^{\pm i\tau^{b}}$, that is, $F^{b}(\Delta_b,\tau^{b}) = F^{b}(\Delta_b,\tau^{b}+\pi)$, as can be verified from Eq.(\ref{eq:holon_free_energy}) in association with Eq.(\ref{eq:holon_energy}).
We note from Eq.(\ref{eq:d_delta_F}) that $\frac{\partial F}{\partial \Delta_b}=0$ at $\Delta_b=0$, thus indicating that there exists a symmetry at $\Delta_b=0$.
However symmetry breaking is found to occur only with the d-wave hole-pairing order but not with the s-wave hole-pairing order.
In the following this will be further elaborated.
For illustration we computed the free energy as a function of holon pair order parameter, $\Delta_b$ by first finding the saddle point values of $\Delta_f = 0.085$ and $\chi=1.63$ with $\delta=0.1$ and $J/t=0.4$.
For the case of d-wave hole(not holon) pairing, $\Delta=<c_{i \uparrow} c_{j \downarrow}>$ which is the composite of the d-wave spinon pairing, $\Delta_f=<f_{i \uparrow} f_{j \downarrow}>$ and the s-wave holon pairing, $\Delta_b=<b_i^\dagger b_j^\dagger>$ in the slave-boson representation we found that there exist minima in the free energy at finite values of $\Delta_b$ at temperature below $T_c$, as is shown in Fig.1.
On the other hand, the s-wave hole pairing did not occur owing to the absence of such minima.
In short, spontaneous symmetry breaking arises as a result of the hole-pairing of the d-wave symmetry, but not the s-wave symmetry at a critical temperature.
In Fig.2 we display computed phase diagrams for various values of $J/t$.
The bose condensation temperature $T_c$ is predicted to increase with $J/t$, indicating that high $T_c$ can be achieved for cuprate materials possessing large values of $J/t$.
The experimentally observed phase diagrams display the bose condensation lines of arch shape with the presence of optimal doping.
The bose condensation temperature is seen to decrease with hole doping rate beyond optimal doping, in agreement with observation\cite{YASUOKA}\cite{SHEN}.

Based on the improved approach to the U(1) slave-boson theory, we found that symmetry breaking can take place as a result of the hole pairing of the d-wave symmetry, but not the s-wave symmetry.
It was shown that a consideration of the supersymmetry Hamiltonian leads to the prediction of the supersymmetry conditions of involving $\Delta_b=\Delta_f=0$ at $T=0 K$, which can occur in the intermediate hole doping region between the antiferromagnetic phase and the superconducting phase.
Finally the decreasing trend of the bose condensation temperature $T_c$ beyond the optimal doping is attributed to the decrease of the pseudogap(spin gap) temperature $T^*$.
This can be readily understood from the expression of the holon pairing term in Eq.(\ref{eq:mf_hamiltonian1}).
Although not shown here, we find that the SU(2) theory\cite{LEE} does not alter the above findings made in the present study.
However the SU(2) theory yields better agreement to the observed phase diagrams of the high $T_c$ cuprates, showing a closer value of optimal doping rate to observation.

One(SHSS) of us acknowledges the generous supports of Korea Ministry of Education(BSRI-98 and 99) and the Center for Molecular Science at Korea Advanced Institute of Science and Technology.
He thanks R. B. Laughlin for helpful discussions on supersymmetry.

%%%%%%%%%%%%%%TWO COLUMN%%%%%%%%%%%%%%%
\begin{minipage}[c]{9cm}
\begin{figure}
\vspace{0cm}
\epsfig{file=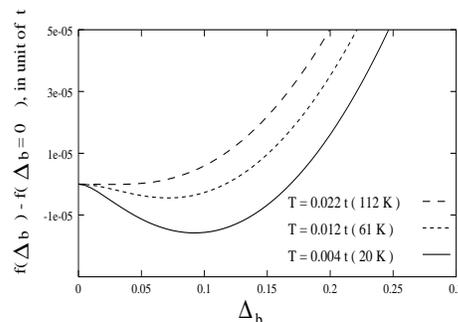,angle=0, height=4cm, width=6cm}
\label{fig:free_energy}
\caption{
Free energy per site as a function of holon pair order parameter.
The computed temperature in the figure is based on $t=0.44eV$[9].
The hole pairing of d-wave symmetry as a composition of the d-wave spinon pair order parameter and the s-wave holon pair order parameter.
}
\end{figure}
 \end{minipage}
 \begin{minipage}[c]{9cm}

 \begin{figure}
 \vspace{0cm}
 \epsfig{file=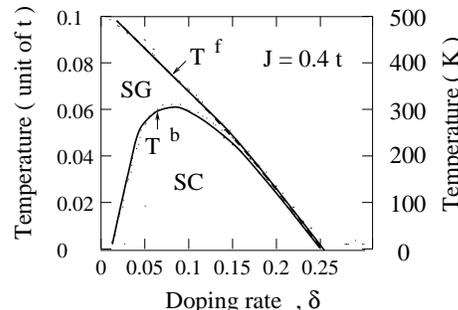, angle=0, height=4cm, width=6cm}
 \label{fig:phase}
 \caption{
Computed phase diagrams with $J/t=0.4$.
$T^f$ denotes the pseudo gap (spinon pair gap) temperature and $T^b$, the holon pair condensation temperature. 
SG stands for the spin gap phase and SC, the superconducting phase.
The scale of temperature in the figure is based on $t=0.44eV$[9].
 }
 \end{figure}
 \end{minipage}
%%%%%%%%%%%%%%%%%%%%%%%%%%%%%%%%%%%%%%%
\end{multicols}


\references
\bibitem{KOTLIAR} G. Kotliar and J. Liu, Phys. Rev. B {\bf 38}, 5142 (1988); references there-in.
\bibitem{FUKUYAMA} Y. Suzumura, Y. Hasegawa and H.  Fukuyama, J. Phys. Soc. Jpn. {\bf 57}, 2768 (1988).
\bibitem{UBBENS} a) M. U. Ubbens and P. A. Lee, Phys. Rev. B {\bf 46}, 8434 (1992); b) M. U. Ubbens and P. A. Lee, Phys. Rev. B {\bf 49}, 6853 (1994); references there-in.
\bibitem{WEN} a) X. G. Wen and P. A. Lee, Phys. Rev. Lett. {\bf 76}, 503 (1996); b) X. G. Wen and P. A. Lee, Phys. Rev. Lett. {\bf 80}, 2193 (1998).
\bibitem{GIMM} T.-H. Gimm, S.-S. Lee, S.-P. Hong and Sung-Ho Suck Salk, Phys. Rev. B, {\bf 60}, 6324 (1999).
\bibitem{HMW} P. C. Hohenberg, Phys. Rev. {\bf 158}, 383 (1967); N.D. Mermin and H.Wagner, Phys Rev. Lett. {\bf 17}, 1133(1966). 
%%
\bibitem{1_DELTA} We readily obtain $\Bigl< b_i b_j b_j^\dagger b_i^\dagger \Bigr> = 1 + \Bigl< b_i^\dagger b_i \Bigr>  + \Bigl< b_j^\dagger b_j \Bigr> +  \Bigl< b_i^\dagger b_i b_j^\dagger b_j \Bigr> \leq 1$. 
Here $b_i b_j b_j^\dagger b_i^\dagger$ is the same as the occupation number operator of the electron pair of charge $-2e$ but not the holon pair of positive charge $+2$ at intersites $i$ and $j$.
It is obvious that the value of $\Bigl<b_i b_j b_j^\dagger b_i^\dagger \Bigr> $ is $1$ for an electron pair occupied at intersites $i$ and $j$ and $0$ for the occupation of the holon(hole)-pair at intersites $i$ and $j$.
It is of note that $\Bigl<  b_i^\dagger b_i \Bigr> = \Bigl<  b_i b_i^\dagger \Bigr>-1 \leq 0 $ with $\Bigl<  b_i b_i^\dagger \Bigr>$, the electron occupation number at site $i$.
We then write $\Bigl<  b_i^\dagger b_i \Bigr> =  -\delta$ allowing the uniform hole doping rate, $\delta$.
Insertion of this relation into the first equation leads to $\Bigl< b_i b_j b_j^\dagger b_i^\dagger \Bigr> = (1-\delta)^2$ with its range of $ 0 \leq \Bigl< b_i b_j b_j^\dagger b_i^\dagger \Bigr> \leq 1$.
%%
\bibitem{LAHIRI} A. Lahiri, P. K. Roy and B. Bagchi, Int. J. Mod. Phys. A {\bf 5}, 1383 (1990); references there-in.
\bibitem{HYBERTSEN} M. S. Hybertsen, E. B. Stechel, M. Schluter and D. R. Jennison, Phys. Rev. B {\bf 41}, 11068 (1990).
\bibitem{YASUOKA} H. Yasuoka, Physica C. {\bf 282-287}, 119 (1997); references there-in.
\bibitem{SHEN} A. G. Loeser, Z. -X. Shen, D. S. Dessau, D. S. Marshall, C. H. Park, P. Fournier and A. Kapitulnik, Science {\bf 273}, 325 (1996).
\bibitem{LEE} S.-S. Lee and Sung-Ho Suck Salk, cond-mat/9907226.
\end{document}